\documentclass[aps,prl,twocolumn,superscriptaddress,floatfix]{revtex4}
\usepackage{amsmath}
\usepackage{amsfonts}
\usepackage{amssymb}
\usepackage{graphicx}
\usepackage{cancel}
\usepackage[switch,columnwise]{lineno}
\usepackage{titlesec}
\usepackage{xcolor}
\usepackage{hyperref}
\usepackage[none]{hyphenat}

\makeatletter
\renewcommand{\section}{\@startsection{section}{1}{0mm}
  {-\baselineskip}{0.5\baselineskip}{\bf\leftline}}

\renewcommand{\subsection}{\@startsection{section}{1}{0mm}
  {-\baselineskip}{0.5\baselineskip}{\bf\leftline}}
\makeatother

\begin{document}

\title{Digital Quantum Simulation of Floquet Topological Phases\\
       with a Solid-State Quantum Simulator}

\author{Bing Chen}%
\affiliation{School of Electronic Science and Applied Physics, Hefei University of Technology, Hefei, Anhui 230009, China}%
\affiliation{State Key Laboratory of Quantum Optics and Quantum Optics Devices, Institute of Opto-Electronics, Shanxi University, Taiyuan, Shanxi 030006, China}%
\author{Shuo Li}%
\affiliation{Stanford Institute for Materials and Energy Sciences, Menlo Park, California 94025, United States}%
\author{Xianfei Hou}%
\affiliation{School of Electronic Science and Applied Physics, Hefei University of Technology, Hefei, Anhui 230009, China}%
\author{Feifei Zhou}%
\affiliation{School of Electronic Science and Applied Physics, Hefei University of Technology, Hefei, Anhui 230009, China}%
\author{Peng Qian}%
\affiliation{School of Electronic Science and Applied Physics, Hefei University of Technology, Hefei, Anhui 230009, China}%

\author{Feng Mei}%
\email{meifeng@sxu.edu.cn}
\affiliation{State Key Laboratory of Quantum Optics and Quantum Optics Devices, Institute
of Laser Spectroscopy, Shanxi University, Taiyuan, Shanxi 030006, China}
\affiliation{Collaborative Innovation Center of Extreme Optics, Shanxi
University,Taiyuan, Shanxi 030006, China}

\author{Suotang Jia}%
\affiliation{State Key Laboratory of Quantum Optics and Quantum Optics Devices, Institute
of Laser Spectroscopy, Shanxi University, Taiyuan, Shanxi 030006, China}
\affiliation{Collaborative Innovation Center of Extreme Optics, Shanxi
University,Taiyuan, Shanxi 030006, China}

\author{Nanyang Xu}%
\email{nyxu@hfut.edu.cn}
\affiliation{School of Electronic Science and Applied Physics, Hefei University of Technology, Hefei, Anhui 230009, China}%
\author{Heng Shen}%
\email{heng.shen@physics.ox.ac.uk}
\affiliation{State Key Laboratory of Quantum Optics and Quantum Optics Devices, Institute of Opto-Electronics, Shanxi University, Taiyuan, Shanxi 030006, China}%
\affiliation{Clarendon Laboratory, University of Oxford, Parks Road, Oxford, OX1 3PU, United Kingdom}%
\affiliation{Collaborative Innovation Center of Extreme Optics, Shanxi
University,Taiyuan, Shanxi 030006, China}

\begin{abstract}
Quantum simulator with the ability to harness the dynamics of complex quantum systems has emerged as a promising platform for probing exotic topological phases. Since the flexibility offered by various controllable quantum systems has enabled to gain insight into quantum simulation of such complicated problems, analog quantum simulator has recently shown its feasibility to tackle problems of exploring topological phases. However, digital quantum simulation and detection of topological phases still remain elusive. Here, we develop and experimentally realize the digital quantum simulation of topological phase with a solid-state quantum simulator at room temperature. Distinct from previous works dealing with static topological phases, the topological phases emulated here are Floquet topological phases. Furthermore, we also illustrate the procedure of digitally simulating a quantum quench and observing the nonequilibrium dynamics of Floquet topological phases. By means of quantum quench, the $0$- and $\pi$-energy topological invariants are unambiguously detected through measuring time-averaged spin polarizations. Our experiment opens up a new avenue to digitally simulate and detect Floquet topological phases with fast-developed programmable quantum simulators.
\end{abstract}
\maketitle

\emph{Introduction}.--Floquet systems generally defined by periodically driven or time-dependent Hamiltonians with $\mathcal{H}(t+T)=\mathcal{H}(t)$ for a fixed period $T$, offers new opportunities to observe quantum Floquet matter~\cite{Floquet}. In aspect of topology, one appealing feature of Floquet engineering is the ability to generate Floquet topological phases of matter~\cite{FTP1,FTP2,FTP3} that is inaccessible in static equilibrium systems, thus providing a way to better understand the associated influence such as Floquet Majorana modes \cite{Jiang,Thakurathi,Kundu}, anomalous topological phases with zero Chern number \cite{FTP1,FTP2}, and chiral topological phase with $0$- or $\pi$-energy topological edge states \cite{Delplace,Lago,Fruchart}. Searching topological phase of matter have raised considerable attention in condensed-matter physics, however, realizing Floquet-engineered topological phases in materials remains theoretical.

\begin{figure}
\centering
\includegraphics[width=0.5\textwidth]{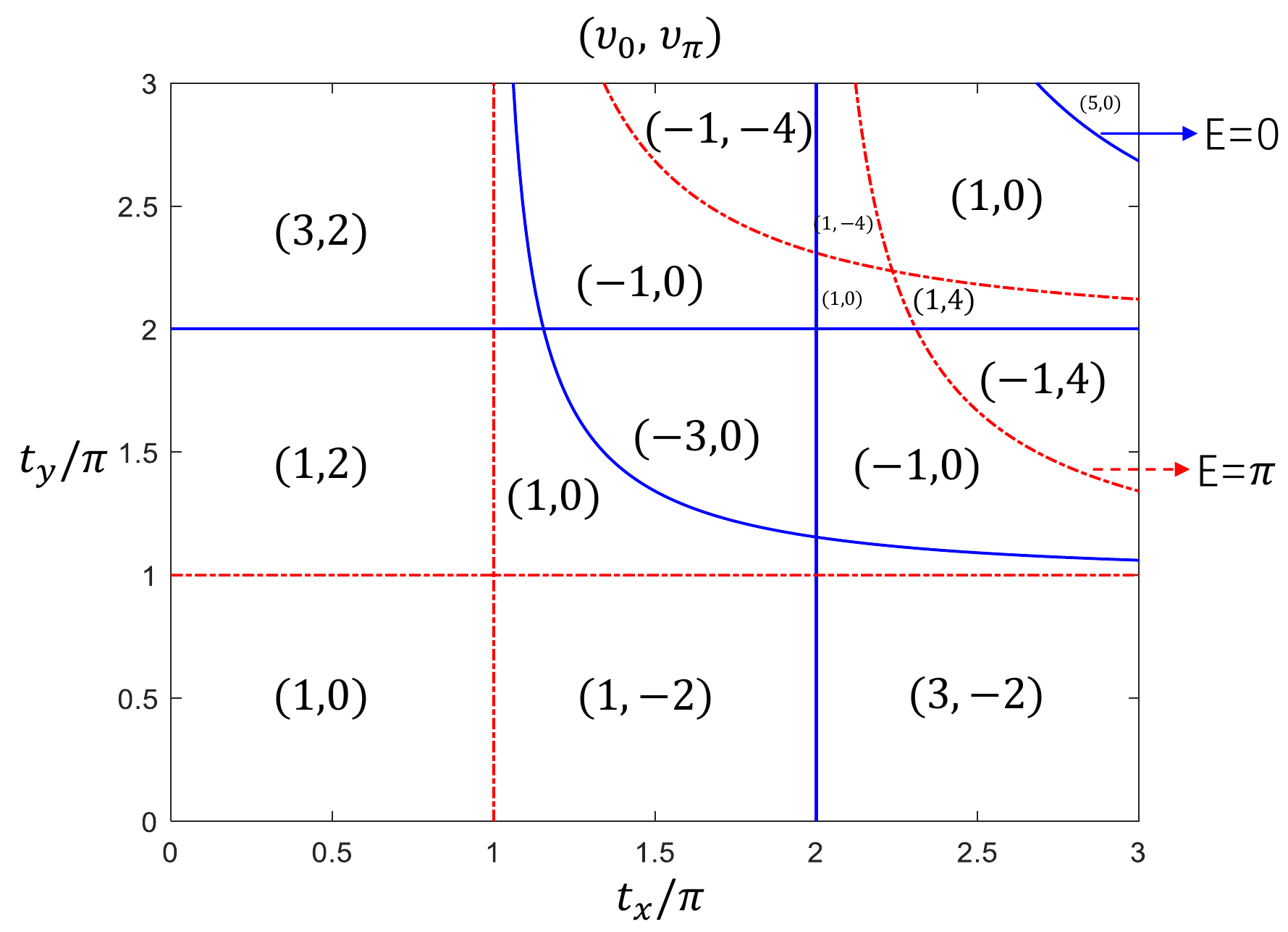}
\caption{The topological winding numbers ($\nu_0$, $\nu_{\pi}$) as a function of $t_x$ and $t_y$. The red and blue solid lines separating different topological phases are corresponding to the gap closings at $E=0$ and $E=\pi$ respectively.  }
\label{Fig1}
\end{figure}


Remarkable advances in quantum simulation \cite{Lloyd,Nori} have revolutionized our understanding of complex system. High degree of controllability enables ultracold atoms \cite{CA1,CA2,Bloch}, trapped ions \cite{Blatt,Monroe}, superconducting circuits \cite{Koch,Martinis} and photonic systems \cite{Walther} to offer the feasibility for tackling problems that are intractable on classical computers. Quantum simulators would not only unveil new results that cannot be otherwise predicted or classically simulated, but they would also allow us to test various models. For instance, analog quantum simulators (AQS) mimic the time evolution of one specific model Hamiltonian, thereby are used to investigate the topological phases and effects with great experimental progresses recently~\cite{TPCA1,TPCA2,TPCA3,Du2016,Du2018,Duan2019,Xu2016,MeiJin2019,Xu2018,MeiSun2019,
Amin2018,Siddiqi2017,Lehnert2014,Wang2016,Roushan2014,
Yin2018,ZhuYu2018,ZhuPan2018,WangYu2019}. As the counterpart of AQS, digital quantum simulation (DQS) \cite{Lloyd,Nori,Blatt2011,DQS3,DQS4,DQS5} encodes the state of the quantum system onto qubits and emulates the time evolution through repeated cycles of qubit rotations (quantum gates) by means of quantum algorithm. Such a circuit-based simulator can, in principle, efficiently simulate any finite-dimensional local Hamiltonian, hence owning the advantage of universality. Although DQS of many-body physics has been intensively studied in various programmable quantum simulators~\cite{Zoller2010,Solano2012a,Solano2012b,Solano2014,
Martinis2015,Wallraff2015,Martinis2016}, digital simulation and detection of topological phases are still less explored \cite{Du2014}.

In this Letter, we report the realization of DQS of Floquet topological phases in a solid-state digital quantum simulator at room temperature \cite{Hanson,Du_npj}. In contrast to the AQS approach simulating static topological phases, it is illustrated here that digital quantum simulator constitutes a natural platform for simulating Floquet topological phases. We further exhibit the DQS of a quantum quench and observe the nonequilibrium dynamics of Floquet topological phases. Most strikingly, after such quench, we show that the signature $0$- and $\pi$-energy topological invariants associated with Floquet topological phases could be detected through measuring the time-averaged spin polarizations, where the experimental observation is in good agreement with theoretical results. We also report the first experimental observation of topological winding number $\nu=5$, much higher than one. Further applications of this protocol could enable studies of high-dimensional and complex Floquet topolocial phases that go beyond the conventional topological systems \cite{Stern,Nayak}.

\begin{figure}
\centering
\includegraphics[width=0.5\textwidth]{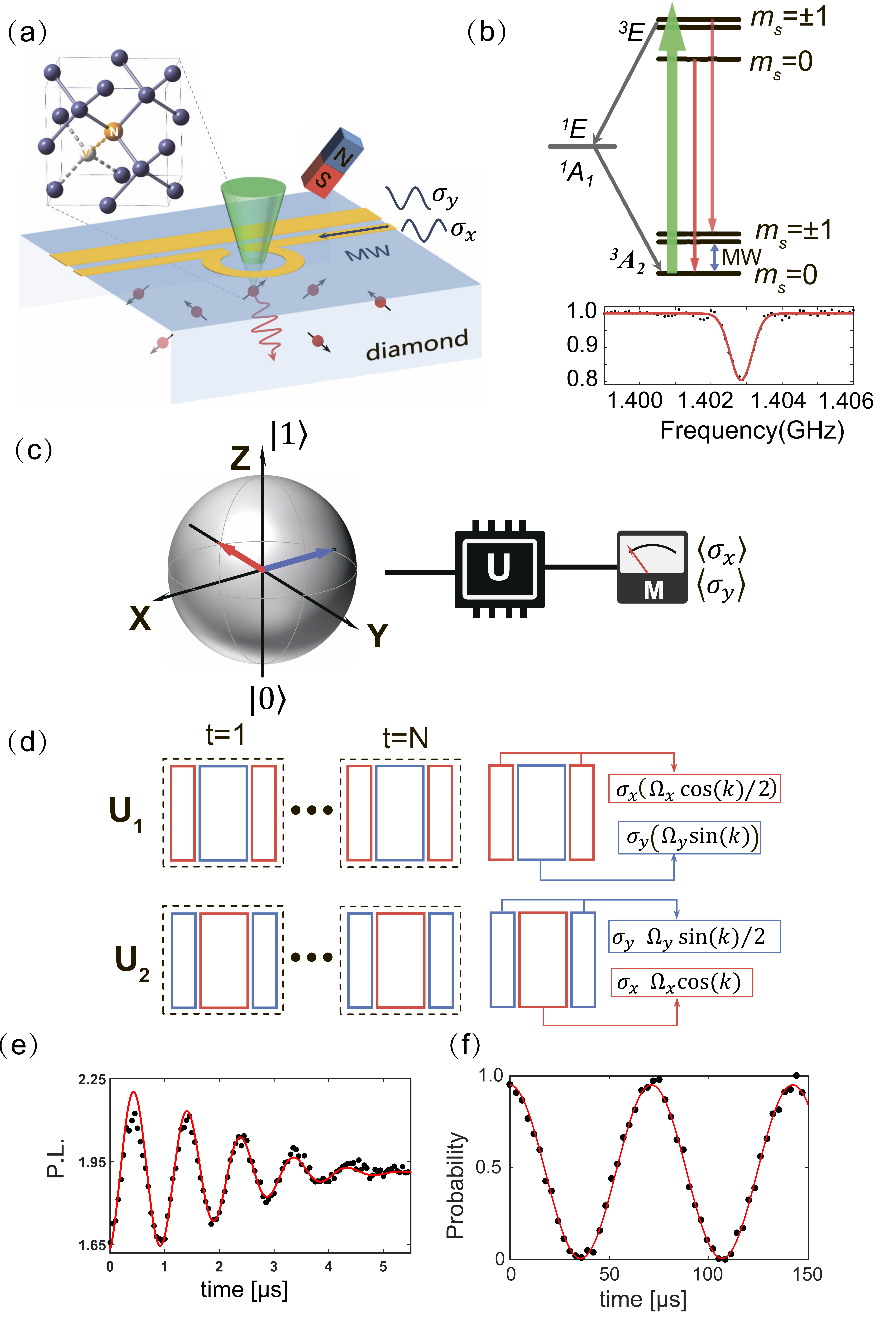}
\caption{(a) Illustration of experiment schematics and atomic structure of the Nitrogen-vacancy (NV) centre in diamond. The NV center in diamond which consists of a substitutional nitrogen atom (N) associated with a vacancy (V) in an adjacent lattice site of the diamond matrix, has $C_{3v}$ symmetry. (b) Scheme of energy levels of the NV center electron spin. Both its ground state ($^3A_2$) and excited state ($^3E$) are spin triplets, and the transition between the two states corresponds to the zero-phonon line (ZPL) at 637nm (1.945 eV). The ground state ($^3A_2$) is a spin triplet with a zero-field splitting of 2.87 GHz between $m_s=0$ and $m_s=\pm 1$ states. The excited states ($^3E$) are governed by spin-orbit and spin-spin interactions, split by 1.43 GHz between $m_s=0$ and $m_s=\pm 1$states. All excited state spin levels (spin quantum number  $m_s=0,\pm 1$) exhibit spontaneous decay by photon emission. The laser and microwave pulse sequence for implementing $U_1$ and $U_2$ are presented in (c) and (d), respectively.}
\label{Fig2}
\end{figure}

\emph{Floquet topological phases}.--In this work, we consider simulating the two-band Floquet topological insulator phases described by the following Floquet Hamiltonian
\begin{equation}
H_F=d_x(k_x)\sigma_x+d_y(k_x)\sigma_y,
\end{equation}
where $d_x$ and $d_y$ are the spin-orbit fields. Specifically, we simulate a one-dimensional periodically driven system formed by two units in one driven period~\cite{Gong2018a,Gong2018b}. Suppose $U$ is the Floquet operator which describes such periodically driven system evolving over one period $T$, i.e.,
\begin{equation}
\hat{U}=e^{-iH_2\frac{T}{2}}e^{-iH_1\frac{T}{2}}.
\end{equation}
where $\hat{H}_1=t_y\sin(k_x)\hat{\sigma}_y$ and $\hat{H}_2=t_x\cos(k_x)\hat{\sigma}_x$. Then, the Floquet Hamiltonian $H_F$ describing the emerged Floquet topological phases is defined as
\begin{equation}
\hat{U}=e^{-i\hat{H}_FT}.
\end{equation}
As we will show, $H_1$ and $H_2$ both can be simulated by apply two microwave pulses on the solid-state NV-center qubit, which further allows us to digitally realize the Floquet operator $U$.

Regarding Floquet topological system, there are two quasienergy gaps centered around $E=0$ and $E=\pi$. The topological features of Floquet topological phases are rooted in these gaps, which are characterized by the topological invariants $\nu_0$ and $\nu_{\pi}$, respectively. According to bulk-edge correspondence associated with topological phases, the values of topological invariants defined in the momentum space count the number of the edge modes defined in the real space. The edge modes are corresponding to the eigenmodes of the real-space lattice Hamiltonian with their densities maximally localized at the edges. In our Floquet topological systems, the value of $\nu_0$ ($\nu_{\pi}$) determines the number of the edge modes with eigenenergy $E=0$ ($E=\pi$).

The topological invariants $\nu_0$ and $\nu_{\pi}$ are defined in terms of a symmetry time framework~\cite{Delplace}, where the starting time point of the Floquet operator $U$ are shifted to two symmetry time points, leading to the following two Floquet operators
\begin{eqnarray}
\hat{U}_1=e^{-i\hat{H}_1\frac{T}{4}}e^{-i\hat{H}_2\frac{T}{2}}e^{-i\hat{H}_1\frac{T}{4}},\nonumber\\
\hat{U}_2=e^{-i\hat{H}_2\frac{T}{4}}e^{-i\hat{H}_1\frac{T}{2}}e^{-i\hat{H}_2\frac{T}{4}}.
\label{U12}
\end{eqnarray}
And both can be further rewritten as in terms of Floquet Hamiltonian
\begin{eqnarray}
\hat{U}_{1}=e^{-i\hat{H}_{F1}T},\,\,\hat{U}_{2}=e^{-i\hat{H}_{F2}T},
\label{HF12}
\end{eqnarray}
where $\hat{H}_{F1,F2}=E\,n_{1,2}\cdot\mathbf{\sigma}$, with $n_{s}=(n_{sx},n_{sy})$ ($s=1,2$)~\cite{n12}, $E=\pm\text{arcos}[\cos\big(t_x\cos(k_x)\big)\cos\big(t_y\sin(k_x)\big)]$, and $\mathbf{\sigma}=(\hat{\sigma}^x,  \hat{\sigma}^y, \hat{\sigma}^z)$ are Pauli spin operators.

Of great interest is the fact that both $\hat{H}_{F1}$ and $\hat{H}_{F2}$ are protected by a chiral symmetry with the chiral operator $\hat{\Gamma}=\hat{\sigma}_z$, supporting chiral topological phases characterized by the following topological winding numbers
\begin{align}
\nu_{s}=\frac{1}{2\pi}\int dk_x(n_{sx}\partial_{k_x}n_{sy}-n_{sy}\partial_{k_x}n_{sx}),\,\, (s=1,\,2.)
\end{align}
The topological invariants $\nu_{0}$ and $\nu_{\pi}$ are defined as
\begin{equation}
\nu_0=\frac{\nu_1+\nu_2}{2},\,\,\nu_{\pi}=\frac{\nu_1-\nu_2}{2}.
\label{v0pi}
\end{equation}

Fig. \ref{Fig1} presents the numerical results of values of the topological invariants $\nu_{0}$ and $\nu_{\pi}$ as a function of $t_x$ and $t_y$. Suprisingly, such a simple periodically driven system has a rich topological phase diagram, and even can support topological phases with topological invariants larger than one~\cite{Gong2018a,Gong2018b}. On the other hand, topological phase transition is known to occur accompanied by a gap closing. In Fig. \ref{Fig1}, we also plot the quasienergy gap closing points at $E=0$ and $E=\pi$. It is shown that the values of the topological invariant $\nu_{0}$ ($\nu_{\pi}$) would change when crossing the gap closing at $E=0$ ($E=\pi$).

\emph{Digital simulation of Floquet topological phases}.--
To digitally simulate the Floquet operator $\hat{U}(T)$ (the time evolution of the Floquet topological Hamiltonian $\hat{H}_F$), we use a negatively charged  nitrogen-vacancy (NV) centre in type-IIa, single-crystal synthetic diamond sample (Element Six).  $m_s=-1$ and $m_s=0$ in ${}^3A_2$ are encoded as spin down $\left | g \right \rangle$ and up $\left | e \right \rangle$ of the electron spin qubit (Fig. \ref{Fig2}). The state of the qubit can be manipulated with microwave pulses ($\omega_{MW}=2\pi\times1404.3 \text{MHz}$), while the spin level $m_s=+1$ remains idle due to large detuning. By applying a laser pulse of 532 nm wavelength with the assistance of intersystem crossing (ISC) transitions, the spin state can be polarized into $m_s=0$ in the ground state. This process can be utilized to initialize and read out the spin state of the NV centre. The fluorescence photons are detected by using the single photon counting module (SPCM). By using a permanent magnet a magnetic field (about 520 G) is applied along the NV axis, the nearby nuclear spins are polarized by optical pumping, improving the coherence time of the electron spin.

The key ingradient in our experiment is to individually engineer the time evolution of the Hamiltonian $\hat{H}_1=t_y\sin(k_x)\hat{\sigma}_y$ and $\hat{H}_2=t_x\cos(k_x)\hat{\sigma}_y$, offered by such a well-controlled solid state quantum simulator \cite{APL}. Both of the Hamiltonians can be emulated by manipulating the electron spin qubit via microwave pulse with $\hat{H}_1=\Omega_y\hat{\sigma}_y,\,\,\hat{H}_2=\Omega_x\hat{\sigma}_x$
and the associated Rabi frequencies $\Omega_x$ and $\Omega_y$. Therefore, by tuning $\Omega_{x}=t_x\cos(k_x)$,  $\Omega_{y}=t_y\sin(k_x)$, Floquet operators depicted in Eq. (\ref{U12}) can be naturally simulated in a digital way [Fig. \ref{Fig2}(c)-(d)]. It is worth mentioning that, this method is genetic and not limited to one-dimensional topological phases studied here. This approach can be also mapped into high-dimensional Brillouin zone and generalized to digital simulation of the Floquet operators associated with high-dimensional Floquet topological phases.

\begin{figure}
\centering
\includegraphics[width=0.5\textwidth]{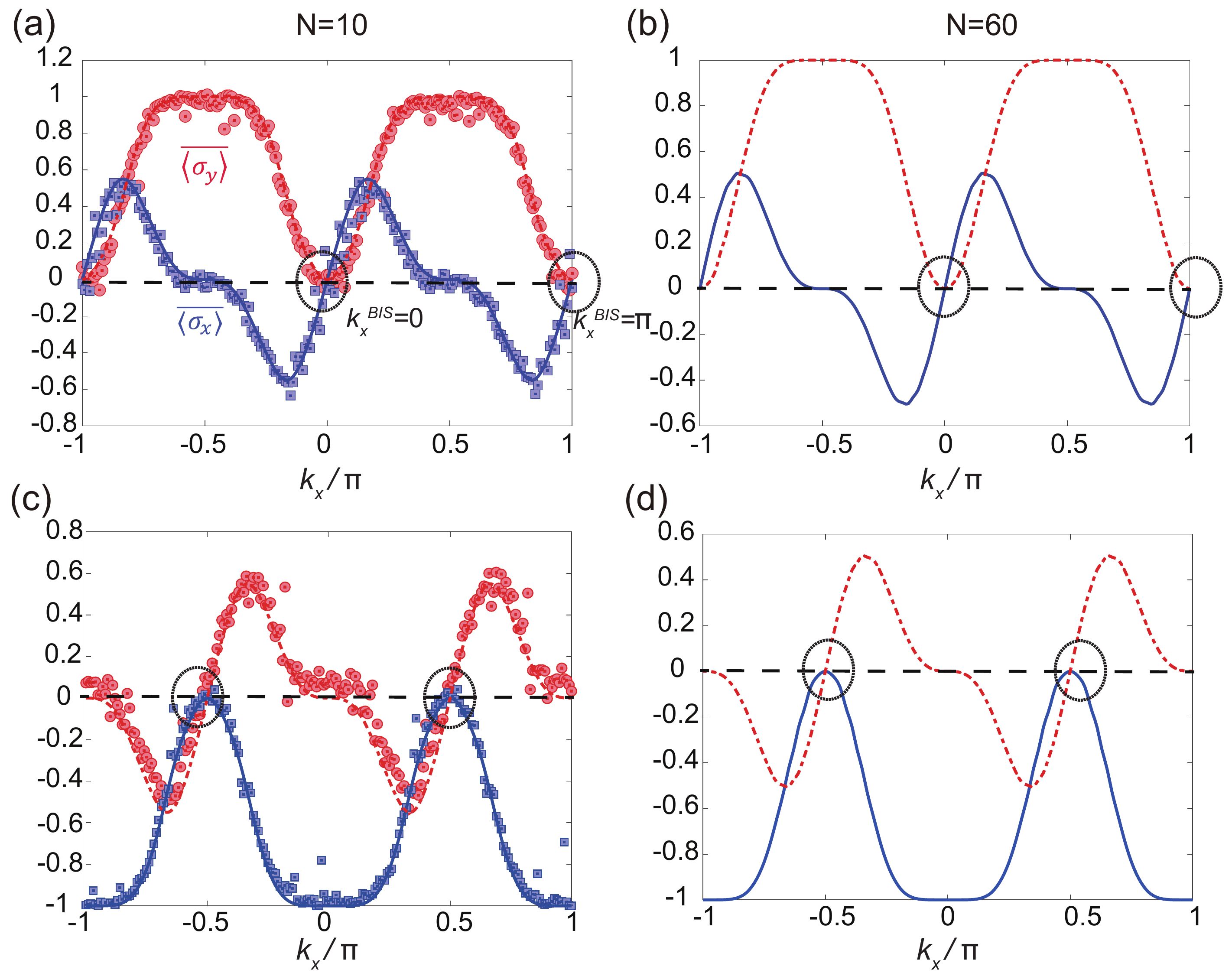}
\caption{The time-averaged spin polarizations $\overline{\langle\hat{\sigma}_{x,y}\rangle}$ as a function of $k_x$ after $N$ times (a,b) $\hat{U}_1$ and (c,d) $\hat{U}_2$. Experimental results with repetition number N=10 are plot in (a) and (c). Solid lines represents theoretical results while red circle and blue square show the experimental values of  $\hat{\sigma}_{y}$ and $\hat{\sigma}_{x}$, respectively. Theoretical results with repetition number N=60 are plot in (b) and (d). The other parameters are $t_x=0.5\pi$ and $t_y=0.5\pi$. Each data point has been averaged $10^6$ repetitions. The error bars account for the statistical error associated with the photon counting.}
\label{Fig3}
\end{figure}

\emph{Digital simulation of quantum quenches and detection of topological invariants}.--
The topological invariants $\nu_{0}$ and $\nu_{\pi}$ featuring the topological properties of the quasienergy gaps centered around $E=0$ and $E=\pi$ are seminal hallmarks of Floquet topological phases~\cite{Delplace}. Recent theoretical study has shown that after a quantum quench the topological invariants associated with static topological phase can be directly measured through the time-averaged spin polarizations on the band inversion surfaces (BISs)~\cite{Liu2018}. We thus proceed to show that such method also can be used to measure the topological invariants associated with Floquet topological phases. Specifically, we digitally perform a quantum quench of Floquet topological phases and employ such quench to measure the topological winding numbers $\nu_{1}$ and $\nu_{2}$, allowing us to detect the topological invariants $\nu_{0}$ and $\nu_{\pi}$ according to Eq. (\ref{v0pi}).

The quantum quench procedure is outlined as follows. First, the initial state of the system is initialized in the ground state of a trivial Floquet topological Hamiltonian $\hat{H}^{i}_F$. Then, $N$ serise of the Floquet operators $\hat{U}_{1,2}$ are digitally performed as shown in Fig.\ref{Fig2} (c) or (d), which digitally simulates the time evolution of a nontrivial Floquet topological Hamiltonian over $N$ periods, i.e.
\begin{equation}
\hat{U}_{1}^N=e^{-iN\hat{H}_{F1}T},\,\,\hat{U}_{2}^N=e^{-iN\hat{H}_{F2}T}.
\end{equation}
As a consequence, a sudden change from a trivial Floquet topological phase to a nontrivial Floquet topological phase is effectively performed, realizing a quantum quench. The initial quantum state will thereby evolve under the final Hamiltonian $\hat{H}_{F1,F2}$.

In our work, two cases in detail are exhibited: {\it (1)} $t_x=0.5\pi$, $t_y=0.5\pi$, with ($\nu_{1}=1$, $\nu_{2}=1$) and ($\nu_{0}=1$, $\nu_{\pi}=0$); {\it (2)} $t_x=2.5\pi$, $t_y=0.5\pi$, with ($\nu_{1}=1$, $\nu_{2}=5$) and ($\nu_{0}=3$, $\nu_{\pi}=-2$). In particular, we implement $y$-direction quench for the Floquet topological Hamiltonian $\hat{H}_{F1}$ and $x$-direction quench for the Floquet topological Hamiltonian $\hat{H}_{F2}$.

For measuring the topological winding number $\nu_1$ in case {\it (1)}, the initial state of the system is prepared into $|\psi(t=0)\rangle=(|g\rangle-i|e\rangle)/\sqrt{2}$, the ground state of a trivial topological Hamiltonian $\hat{H}^i_{F1}=n_{1x}\hat{\sigma}_x+(m_y+n_{1y})\hat{\sigma}_y$ with $m_y>>1$. After that, we repeat $\hat{U}_1$ operation $N$ times. Consequently, the dynamics of the initial state $|\psi(t=0)\rangle$ is govern by a nontrivial topological Hamiltonian $\hat{H}_{F1}$, fufilling a quantum quench from $m_y>>1$ to $m_y=0$. In this $y$-direction quench process, the BIS appears when $n_{1y}=0$~\cite{Liu2018}, which yields $k^{BIS}_{x}=0,\pi$. After the quantum quench, we measure the time evolution of the spin polarization $\langle\hat{\sigma}_{x,y}\rangle$ for each $k_x$ in the Brilliouin zone, from which we extract the time-averaged spin polarization
$\overline{\langle\hat{\sigma}_{x,y}\rangle}=\frac{1}{N}\sum^N_{t=1}\langle\hat{\sigma}_{x,y}\rangle_t$, with the time-resolved spin polarization $\langle\hat{\sigma}_{x,y}\rangle_t=\langle\psi(t=0)|(\hat{U}^{-1}_1)^t\hat{\sigma}_{x,y}(\hat{U}_1)^t|\psi(t=0)\rangle$.
When $N$ is very large, the BISs $k^{BIS}_{x}$ and the topological winding number $\nu_1$ both can be directly measured through $\overline{\langle\hat{\sigma}_{x,y}\rangle}$~\cite{Liu2018}, i.e.,
\begin{align}
&\overline{\langle\hat{\sigma}_{y}(k^{BIS}_{x})\rangle}=0,\nonumber \\
&\nu_1=\frac{1}{2}(g_x(k^{BIS}_{x}=\pi)-g_x(k^{BIS}_{x}=0)),
\label{quench}
\end{align}
where $g_x(k^{BIS}_{x})=-\partial_{k_{\perp}}\overline{\langle\hat{\sigma}_{x}\rangle}$ is related to the slope of the time-averaged spin polarization $\overline{\langle\hat{\sigma}_{x}\rangle}$ at BISs, with $k_{\perp}$ denoting the momentum perpendicular to BIS and points from $n_{1y}<0$ to $n_{1y}>0$. The theoretical results for $N=60$ are presented in Fig. \ref{Fig3}(b), showing that $\overline{\langle\hat{\sigma}_{y}(k^{BIS}_{x}=0,\pi)\rangle}=0$ and $g_x(k^{BIS}_{x}=\pi)=-g_x(k^{BIS}_{x}=0)=1$. In practice, we find that all the above results hold true even for $N=10$ experimentally. Fig. \ref{Fig3}(a) shows the experimental measured results of $\overline{\langle\hat{\sigma}_{x,y}\rangle}$ in good agreement with the theoretical results. Based on these results, it is found that the time-averaged spin polarization $\overline{\langle\hat{\sigma}_{y}\rangle}$ is zero when $k_{x}=0,\pi$, which allows us to clearly identify the BISs. Note that $n_{1y}>0$ when $k_x\in(0,\pi)$, otherwise $n_{1y}<0$. Therefore, the slope of the time-averaged spin polarization $\overline{\langle\hat{\sigma}_{x}\rangle}$ at $k^{BIS}_{x}=0$ ($k^{BIS}_{x}=\pi$) is recognized as 1 (-1), yielding $g_x(k^{BIS}_{x}=\pi)=-g_x(k^{BIS}_{x}=0)=1$. According to Eq. (\ref{quench}), we can determine the topological winding number value as $\nu_1=1$.

For measuring the topological winding number $\nu_2$ in case {\it (1)}, the initial state of the system is prepared into $|\psi(t=0)\rangle=(|g\rangle-|e\rangle)/\sqrt{2}$, the ground state of a trivial topological Hamiltonian $\hat{H}^i_{F2}=(m_x+n_{2x})\hat{\sigma}_x+n_{2y}\hat{\sigma}_y$ with $m_x>>1$. Subsequently, a $x$-direction quantum quench from $m_x>>1$ to $m_x=0$ is implemented by repeating $\hat{U}_2$ $N$ times and thus causing the time evolution of the initial state $|\psi(t=0)\rangle$ to be govern by a nontrivial topological Hamiltonian $\hat{H}_{F2}$. The BIS appears when $n_{2x}=0$~\cite{Liu2018}, which gives $k^{BIS}_{x}=\pm0.5\pi$. The BISs and the topological winding number $\nu_2$ are measured through~\cite{Liu2018}
\begin{align}
 &\overline{\langle\hat{\sigma}_{x}(k^{BIS}_{x})\rangle}=0, \nonumber \\ &\nu_2=\frac{1}{2}(g_y(k^{BIS}_{x}=0.5\pi)-g_y(k^{BIS}_{x}=-0.5\pi)),
\end{align}
where $g_y(k^{BIS}_{x})=-\partial_{k_{\perp}}\overline{\langle\hat{\sigma}_{y}\rangle}$ is related to the slope of the time-averaged spin polarization $\overline{\langle\hat{\sigma}_{y}\rangle}$ at the BISs, and with $k_{\perp}$ denoting the momentum perpendicular to BIS and points from $n_{2x}<0$ to $n_{2x}>0$. As theoretically presented in Fig. \ref{Fig3}(d) for $N=60$,  $\overline{\langle\hat{\sigma}_{x}(k^{BIS}_{x}=0,\pi)\rangle}=0$ and $g_y(k^{BIS}_{x}=0.5\pi)=-g_y(k^{BIS}_{x}=-0.5\pi)=1$.  The corresponding experimental results for $N=10$ are shown in Fig. \ref{Fig3}(c) and agree well with theoretical results. From these experimental results, we can unambiguously conclude $\overline{\langle\hat{\sigma}_{x}(k_x=\pm0.5\pi)\rangle}=0$ and $g_y(k^{BIS}_{x}=0.5\pi)=-g_y(k^{BIS}_{x}=-0.5\pi)=1$, identifying the location of BISs and the value of the topological winding number $\nu_2=1$. Hence, according to Eq. (\ref{v0pi}), the $0$- and $\pi$-energy topological invariants are measured as $\nu_0=1$ and $\nu_{\pi}=0$.

\begin{figure}
\centering
\includegraphics[width=0.5\textwidth]{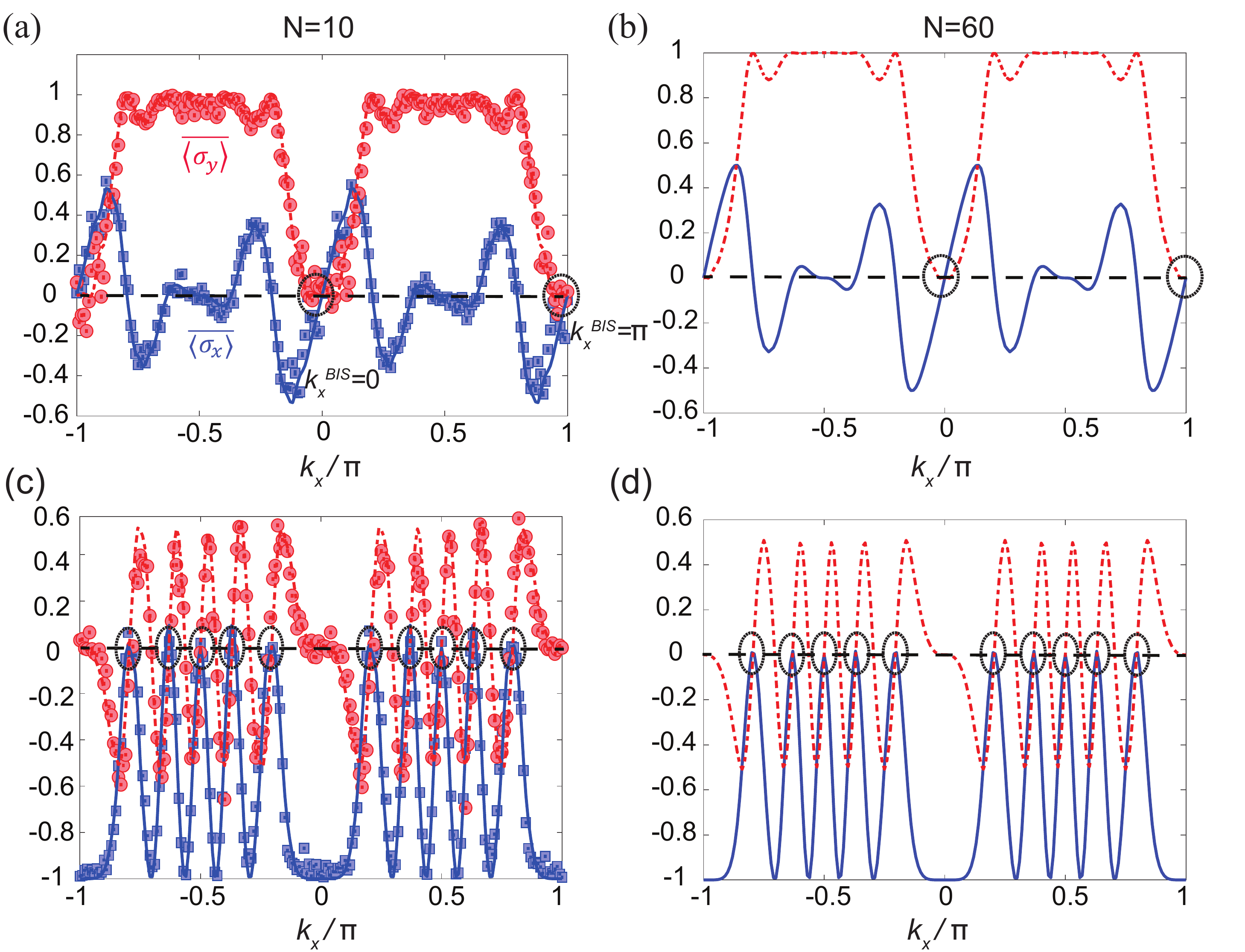}
\caption{The time-averaged spin polarizations $\overline{\langle\hat{\sigma}_{x,y}\rangle}$ as a function of $k_x$ after $N$ times (a,b) $\hat{U}_1$ and (c,d) $\hat{U}_2$. Experimental results with repetition number N=10 are plot in (a) and (c). Solid lines represents theoretical results while red circle and blue square show the experimental values of  $\hat{\sigma}_y$ and $\hat{\sigma}_x$, respectively.
Theoretical results with repetition number N=60 are plot in (b) and (d). The other parameters are $t_x=2.5\pi$ and $t_y=0.5\pi$. Each data point has been averaged $10^6$ repetitions. The error bars account for the statistical error associated with the photon counting.}
\label{Fig4}
\end{figure}

Regarding case {\it (2)}, the same procedure is applied to extract the topological winding number $\nu_1$ and $\nu_2$. The experimental results on the time-averaged spin polarizations after $N=10$ times $\hat{U}_1$ and $\hat{U}_2$ are shown in Fig. \ref{Fig4}(a) and (c) respectively, which agree well with the theoretical results. Fig. \ref{Fig4}(a) illustrates that the BIS appears at $k^{BIS}_{x}=0,\pi$ in which $\overline{\langle\sigma_{y}\rangle}=0$ and $g_x(k^{BIS}_{x}=\pi)=-g_x(k^{BIS}_{x}=0)=1$. The value of the topological winding number is measured through $\nu_1=\frac{1}{2}(g_x(k^{BIS}_{x}=\pi)-g_x(k^{BIS}_{x}=0))=1$.
Fig. \ref{Fig4}(c) shows that the BIS appears at $k^{BIS}_{x}=\pm0.5\pi,\pm\text{arcos}(\pm0.4),\pm\text{arcos}(\pm0.8)$. At such points, $\overline{\langle\hat{\sigma}_{x}\rangle}$ is not strictly zero but still maximal and approaching zero. As suggested by the theoretical results shown in Fig. \ref{Fig4}(d) for $N=60$, the slight difference from zero is a result of the fact that the number $N$ we chose in the experiment is not larger enough. This does not affect the measurement of the slopes of the time-averaged spin polarization $\overline{\langle\sigma_{y}\rangle}$ at the BISs, i.e., $g_y(k^{BIS}_{x}\in k_+)=-g_y(k^{BIS}_{x}\in k_{-})=1$, where $k_{+}=0.5\pi,\text{arcos}(\pm0.8),-\text{arcos}(\pm0.4)$ and $k_{-}=-0.5\pi,-\text{arcos}(\pm0.8),\text{arcos}(\pm0.4)$. The topological winding number $\nu_2$ is measured through $\nu_2=\frac{1}{2}(\sum_{k^{BIS}_{x}\in k_{+}}g_y(k^{BIS}_{x})
-\sum_{k^{BIS}_{x}\in k_{-}}g_y(k^{BIS}_{x}))$, giving the value of the topological winding number as $\nu_{2}=5$. Substituting the above values into Eq. (\ref{v0pi}), the $0$- and $\pi$-energy topological invariants are determined as $\nu_0=3$ and $\nu_{\pi}=-2$.

We emphasize that $g_{x,y}(k^{BIS}_{x})$ are related to the slopes of the time-averaged spin polarizations $\overline{\langle\sigma_{x,y}\rangle}$ at the BISs and quite robust to the experimental imperfections. This feature manifests topological protection and enables accurate measurements of the topological invariants of the Floquet topological phases.

\emph{Summary and Outlook}-- In summary, we have reported the digital simulation and detection of Floquet topological phases with a solid-state quantum simulator. In order to measure Floquet topological invariants, quantum quenches of Floquet topological phases are digitally simulated. The method developed here can be directly applied to other well-developed platforms of quantum simulator, such as superconduction circuits and trapped ions. Our work opens a door for DQS of topological phases with programmable quantum simulators, including high-dimensional Floquet topological insulators~\cite{FTP1,FTP2,FTP3,3DFTP}, Floquet $Z_2$ topological phases~\cite{Z2FTP} and Floquet Hopf insulators~\cite{FHI} that are hard to be engineered in other topological systems. This also paves the way for DQS of nonequilibrium topological phases~\cite{Hu2016,Refael2016,Mueller2016}, where quantum quenches could be digitally simulated.

\emph{Acknowledgements}--The authors acknowledge the financial support from the National Key Research and Development Program of China (2018YFA0306600, and 2018YFF01012500), Natural National Science Foundation of China (NSFC)(11604069, 11904070), the Program of State Key Laboratory of Quantum Optics and Quantum Optics Devices (No.KF201802) and the Fundamental Research Funds for the Central Universities (PA2019GDQT0023).
F. M. was supported by the National Key Research and Development Program of China (2017YFA0304203), Natural National Science Foundation of China (NNSFC) (12074234), Fund for Shanxi 1331 Project Key Subjects Construction, PCSIRT (No. IRT-17R70), and 111 Project (D18001). H. Shen acknowledges the financial support from the Royal Society Newton International Fellowship (NF170876) of United Kingdom.


\begin{thebibliography}{9}\label{sec:TeXbooks}%

\bibitem{Floquet}
R. Moessner and S. L. Sondhi, Equilibration and order in quantum Floquet matter, \emph{Nat. Phys.} \textbf{13}, 424 (2017).

\bibitem{FTP1}
T. Kitagawa, E. Berg, M. Rudner, E. Demler, Topological characterization of periodically driven quantum systems, \emph{Phys. Rev. B} \textbf{82}, 235114 (2010).

\bibitem{FTP2}
M. S. Rudner, N. H. Lindner, E. Berg, and M. Levin, Anomalous edge states and the bulk-edge correspondence for periodically driven two-dimensional systems, \emph{Phys. Rev. X} \textbf{3}, 031005 (2013).

\bibitem{FTP3}
N. H. Lindner, G. Refael, and V. Galitski, Floquet topological insulator in semiconductor quantum wells, \emph{Nat. Phys.} \textbf{7}, 490 (2011).


\bibitem{Jiang}
L. Jiang, \emph{et al.}, Majorana Fermions in Equilibrium and in Driven Cold-Atom Quantum Wires, \emph{Phys. Rev. Lett.} \textbf{106}, 220402 (2011).
%
\bibitem{Thakurathi}
M. Thakurathi, A. A. Patel, D. Sen, and A. Dutta, Floquet generation of Majorana end modes and topological invariants, \emph{Phys. Rev. B} \textbf{88}, 155133 (2013).
%
\bibitem{Kundu}
A. Kundu and B. Seradjeh, Transport Signatures of Floquet Majorana Fermions in Driven Topological Superconductors, \emph{Phys. Rev. Lett.} \textbf{111}, 136402 (2013).
%
%
\bibitem{Delplace}
J. K. Asb\'oth, B. Tarasinski, and P. Delplace, Chiral symmetry and bulk-boundary correspondence in periodically driven one-dimensional systems, \emph{Phys. Rev. B} \textbf{90}, 125143 (2014).
%
\bibitem{Lago}
V. Dal Lago, M. Atala, and L. E. F. Foa Torres, Floquet topological transitions in a driven one-dimensional topological insulator, \emph{Phys. Rev. A} \textbf{92}, 023624 (2015).
%
\bibitem{Fruchart}
M. Fruchart, Complex classes of periodically driven topological lattice systems, \emph{Phys. Rev. B} \textbf{93}, 115429 (2016).
%
\bibitem{Lloyd}
S. Lloyd, Universal quantum simulators, \emph{Science} \textbf{273}, 1073-1078 (1996).
%
\bibitem{Nori}
I. Buluta, F. Nori, Quantum Simulators, \emph{Science} \textbf{326}, 108-111 (2009).

\bibitem{CA1} I. Bloch, J. Dalibard, and W. Zwerger, Many-body physics with ultracold gases, \emph{Rev. Mod. Phys.} \textbf{80}, 885 (2008).

\bibitem{CA2} M. Lewenstein, \emph{et al.}, Ultracold atomic gases in optical lattices: mimicking condensed matter physics and beyond, \emph{Adv. Phys.} \textbf{56}, 243 (2007).

\bibitem{Bloch}
C. Gross and I. Bloch, Quantum simulaitons with ultracold atoms in optical lattices, \emph{Science} \textbf{357}, 995-1001 (2017).
%
%
\bibitem{Blatt}
R. Blatt and C. F. Roos, Quantum simulations with trapped ions, \emph{Nat. Phys.} \textbf{8}, 277-284 (2012).
%
\bibitem{Monroe}
K. Kim, M.-S. Chang, S. Korenblit, R. Islam, E. E. Edwards, J. K. Freericks, G.-D. Lin, L.-M. Duan, and C. Monroe, Quantum Simulation of Frustrated Ising Spins with Trapped Ions, \emph{Nature} \textbf{465}, 590 (2010).
%
\bibitem{Koch}
A. A. Houck, H. E. T\"ureci and J. Koch, On-chip quantum simulation with superconducting circuits. \emph{Nat. Phys.} \textbf{12}, 292-299 (2012).
%
\bibitem{Martinis}
F. Arute \emph{et al.}, Quantum supremacy using a programmable superconducting processor, \emph{Nature} \textbf{574}, 505-510 (2019).
%
\bibitem{Walther} A. Aspuru-Guzik and P. Walther, Photonic quantum simulators, \emph{Nat. Phys.} \textbf{12}, 285-291 (2012).

\bibitem{TPCA1}
N. Goldman, J. C. Budich, and P. Zoller, Topological quantum matter with ultracold gases in optical lattices,
\emph{Nat. Phys.} \textbf{12}, 639 (2016).

\bibitem{TPCA2}
D.-W. Zhang, Y.-Q. Zhu, Y. X. Zhao, H. Yan, and S.-L. Zhu, Topological quantum matter with cold atoms, \emph{Adv. Phys.}
\textbf{67}, 253 (2019).
%
\bibitem{TPCA3}
N. R. Cooper, J. Dalibard, and I. B. Spielman, Topological bands for ultracold atoms, \emph{Rev. Mod. Phys.} \textbf{91}, 015005 (2019).
%
\bibitem{Du2016}
F. Kong, \emph{et al.}, Direct Measurement of Topological Numbers with Spins in Diamond, \emph{Phys. Rev. Lett.} \textbf{117}, 060503 (2016).

\bibitem{Du2018}
W. Ma, \emph{et al.}, Experimental Observation of a Generalized Thouless Pump with a Single Spin, \emph{Phys. Rev. Lett.} \textbf{120}, 120501 (2018).

\bibitem{Duan2019}
W. Lian, \emph{et al.}, Machine learning topological phases with a solid-state quantum simulator, \emph{Phys. Rev. Lett.}
\textbf{122}, 210503 (2019).

\bibitem{Xu2016}
J. S. Xu, K. Sun, Y. J. Han, C. F. Li, J. K. Pachos, and G. C. Guo, Simulating the exchange of Majorana zero modes with a photonic system, \emph{Nat. Commun.} \textbf{7}, 13194 (2016).

\bibitem{MeiJin2019}
Y. Wang, Y. H. Lu, F. Mei, J. Gao, Z. M. Li, H. Tang, S. L. Zhu, S. Jia, and X. M. Jin, Direct Observation of Topology from Single-photon Dynamics on a Photonic Chip,  \emph{Phys. Rev. Lett.} \textbf{122}, 193903 (2019).

\bibitem{Xu2018} J. S. Xu, K. Sun, J. K. Pachos, Y. J. Han, C. F. Li, and G. C. Guo, Photonic implementation of Majorana-based Berry phases, \emph{Sci. Adv.} \textbf{4}, 6533 (2018).


\bibitem{MeiSun2019}
W. Cai, \emph{et al.}, Observation of topological magnon insulator states in a superconducting circuit, \emph{Phys. Rev. Lett.} \textbf{123}, 080501 (2019).

\bibitem{Amin2018}
A. D. King,\emph{et al.}, Observation of topological phenomena in a programmable lattice of 1,800 qubits, \emph{Nature (London)} \textbf{560}, 456 (2018).
%
\bibitem{Siddiqi2017}
E. Flurin, V. V. Ramasesh, S. Hacohen-Gourgy, L. S. Martin, N. Y. Yao, and I. Siddiqi, Observing Topological Invariants Using Quantum Walks in Superconducting Circuits, \emph{Phys. Rev. X} \textbf{7}, 031023 (2017).

\bibitem{Lehnert2014}
M. D. Schroer, M. H. Kolodrubetz, W. F. Kindel, M. Sandberg, J. Gao, M. R. Vissers, D. P. Pappas, A. Polkovnikov, and K. W. Lehnert, Measuring a Topological Transition in an Artificial Spin-$1/2$ System, \emph{Phys. Rev. Lett.} \textbf{113}, 050402
(2014).

\bibitem{Roushan2014}
P. Roushan, \emph{et al.}, Observation of topological transitions in interacting quantum circuits, \emph{Nature (London)} \textbf{515}, 241 (2014).

\bibitem{Wang2016}
Y. P. Zhong, \emph{et al.}, Emulating Anyonic Fractional Statistical Behavior in a Superconducting Quantum Circuit, \emph{Phys. Rev. Lett.} \textbf{117}, 110501 (2016).

\bibitem{Yin2018}
T. Wang, Z. Zhang, L. Xiang, Z. Gong, J. Wu, and Y. Yin, Simulating a topological transition in a superconducting phase qubit by fast adiabatic trajectories, \emph{Sci. China Phys. Mech. Astron.} \textbf{61}, 047411 (2018).

\bibitem{ZhuYu2018}
X. Tan, D.-W. Zhang, Q. Liu, G. Xue, H.-F. Yu, Y. -Q. Zhu, H. Yan, S.-L. Zhu, and Y. Yu, Topological Maxwell metal
bands in a superconducting qutrit, \emph{Phys. Rev. Lett.} \textbf{120}, 130503 (2018).

\bibitem{ZhuPan2018}
C. Song, \emph{et al.}, Demonstration of topological robustness of anyonic braiding statistics with a superconducting quantum circuit, \emph{Phys. Rev. Lett.} \textbf{121}, 030502 (2018).

\bibitem{WangYu2019}
X. Tan, Y. X. Zhao, Q. Liu, G. Xue, H. F. Yu, Z. D. Wang, and Y. Yu, Simulation and manipulation of tunable Weyl-semimetal bands using superconducting quantum circuits, \emph{Phys. Rev. Lett.} \textbf{122}, 010501 (2019).

\bibitem{Blatt2011}
B. P. Lanyon, \emph{et al.}, Universal digital quantum simulation with trapped ions, \emph{Science} \textbf{334}, \textbf{57} (2011).

\bibitem{DQS3}
M. M\"{u}ller, S. Diehla, G. Pupillo, P. Zoller, Engineered Open Systems and Quantum Simulations with Atoms and Ions, \emph{Advances In Atomic, Molecular, and Optical Physics} \textbf{61}, 1 (2012).

\bibitem{DQS4}
I. M. Georgescu, S. Ashhab, and F. Nori, Quantum simulation, \emph{Rev. Mod. Phys.} \textbf{86}, 153 (2014).

\bibitem{DQS5}
L. Lamata, A. Parra-Rodriguez, M. Sanz, and E. Solano, Digital-analog quantum simulations with superconducting circuits, \emph{Advances in Physics: X} \textbf{3}, 1457981 (2018).
%
\bibitem{Zoller2010}
H. Weimer, M. M\"{u}ller, I. Lesanovsky, P. Zoller and H. Peter B\"{u}chler, A Rydberg quantum simulator, \emph{Nat. Phys.} \textbf{6}, 382 (2010).

\bibitem{Solano2012a}
A. Mezzacapo, J. Casanova, L. Lamata, and E. Solano, Digital Quantum Simulation of the Holstein Model in Trapped Ions, \emph{Phys. Rev. Lett.} \textbf{109}, 200501 (2012).

\bibitem{Solano2012b}
J. Casanova, A. Mezzacapo, L. Lamata, and E. Solano, Quantum Simulation of Interacting Fermion Lattice Models in Trapped Ions, \emph{Phys. Rev. Lett.} \textbf{108}, 190502 (2012).

\bibitem{Solano2014}
U. L. Heras, A. Mezzacapo, L. Lamata, S. Filipp, A. Wallraff, and E. Solano, Digital Quantum Simulation of Spin Systems in Superconducting Circuits, \emph{Phys. Rev. Lett.} \textbf{112}, 200501 (2014).

\bibitem{Martinis2015}
R. Barends, \emph{et al.}, Digital quantum simulation of fermionic models with a superconducting circuit, \emph{Nat. Commun.} \textbf{6}, 7654 (2015).

\bibitem{Wallraff2015}
Y. Salath\'{e}, \emph{et al.}, Digital quantum simulation of spin models with circuit quantum electrodynamics, \emph{Phys. Rev. X} \textbf{5}, 021027 (2015).

\bibitem{Martinis2016}
R. Barends, \emph{et al.}, Digitized adiabatic quantum computing with a superconducting circuit, \emph{Nature (London)} \textbf{534}, 222 (2016).

\bibitem{Du2014}
C. Ju, C. Lei, X. Xu, D. Culcer, Z. Zhang, and J. Du, NV-center-based digital quantum simulation of a quantum phase transition in topological insulators, \emph{Phys. Rev. B} 89, 045432 (2014).
%

\bibitem{Hanson}
T. van der Sar \emph{et al.}, Decoherence-protected quantum gates for a hybrid solid-state spin register, \emph{Nature} \textbf{484}, 82-86 (2012).
%
\bibitem{Stern}
A. Stern, N. H. Lindner, Topological quantum computation-from basic concepts to first experiments, \emph{Science} 339, 1179 (2013).
%
\bibitem{Nayak}
C. Nayak, S. H. Simon, A. Stern, M. Freedman, and S. D. Sarma, Non-Abelian anyons and topological quantum
computation, \emph{Rev. Mod. Phys.} \textbf{80} 1083 (2008).
%
\bibitem{Du_npj}
Y. Wu, Y. Wang, X. Qin, X. Rong, and J. Du, A programmable two-qubit solid-state quantum processor under ambient conditions, \emph{npj Quantum Inf.} \textbf{5}, 9 (2019).

\bibitem{Gong2018a}
L. Zhou and J. Gong, Floquet topological phases in a spin-1/2 double kicked rotor, \emph{Phys. Rev. A} \textbf{97}, 063603 (2018).

\bibitem{Gong2018b}
L. Zhou and J. Gong, Non-Hermitian Floquet topological phases with arbitrarily many real-quasienergy edge states, \emph{Phys. Rev. B} \textbf{98}, 205417 (2018).

%
%
\bibitem{n12}
$n_{1x}=\sin\big(t_x\cos(k_x)\big)\cos\big(t_y\sin(k_x)\big)$,
$n_{1y}=\sin\big(t_y\sin(k_x)\big)$,
$n_{2x}=\sin\big(t_x\cos(k_x)\big)$,
$n_{2y}=\cos\big(t_x\cos(k_x)\big)\sin\big(t_y\sin(k_x)\big)$.

\bibitem{APL}
B. Chen \emph{et al.}, Quantum state tomography of a single electron spin in diamond with Wigner function reconstruction, \emph{Appl. Phys. Lett.} \textbf{116}, 041102 (2019).

\bibitem{Liu2018}
L. Zhang, L. Zhang, S. Niu, and X.-J. Liu, Dynamical classification of topological quantum phases, \emph{Science Bulletin} \textbf{63}, 1385 (2018).

\bibitem{3DFTP}
N. H. Lindner, D. L. Bergman, G. Refael, and V. Galitski, Topological Floquet spectrum in three dimensions via a two-photon resonance, \emph{Phys. Rev. B} \textbf{87}, 235131 (2013).

\bibitem{Z2FTP}
D. Carpentier, P. Delplace, M. Fruchart, and K. Gawedzki, Topological Index for Periodically Driven Time-Reversal Invariant 2D Systems, \emph{Phys. Rev. Lett.} \textbf{114}, 106806 (2015).

\bibitem{FHI}
T. Schuster, S. Gazit, J. E. Moore, and N. Y. Yao, Floquet Hopf Insulators, \emph{Phys. Rev. Lett.} \textbf{123}, 266803 (2019).

\bibitem{Hu2016}
Y. Hu, P. Zoller, and J. C. Budich, Dynamical Buildup of a Quantized Hall Response from Nontopological States, \emph{Phys. Rev. Lett.} \textbf{117}, 126803 (2016).

\bibitem{Refael2016}
J. H. Wilson, J. C. W. Song, and G. Refael, Remnant geometric Hall response in a quantum quench, \emph{Phys. Rev. Lett.} \textbf{117}, 235302 (2016).

\bibitem{Mueller2016}
F. N. Unal, E. J. Mueller, and M. O. Oktel, Nonequilibrium fractional Hall response after a topological quench, \emph{Phys. Rev. A} \textbf{94}, 053604 (2016).

\end{thebibliography}
\end{document}